\documentstyle[titlepage,12pt]{article}
\begin{document}
\newfont{\bg}{cmr10 scaled\magstep3}
%
\newcommand{\gsimeq}
{\hbox{ \raise3pt\hbox to 0pt{$>$}\raise-3pt\hbox{$\sim$} }}
\newcommand{\lsimeq}
{\hbox{ \raise3pt\hbox to 0pt{$<$}\raise-3pt\hbox{$\sim$} }}


%
%


%
%

\newcommand{\plb}[3]{Phys. Lett. {\bf B#1}, (#2), #3} 
\newcommand{\prl}[3]{Phys. Rev. Lett. {\bf #1}, (#2), #3}
\newcommand{\prd}[3]{Phys. Rev. {\bf D#1}, (#2), #3}
\newcommand{\npb}[3]{Nucl. Phys. {\bf B#1}, (#2), #3}
\newcommand{\npbps}[3]{Nucl. Phys. {\bf B}(Proc. Suppl.) {\bf #1}, (#2), #3}
\newcommand{\prog}[3]{Prog. Theor. Phys. {\bf #1}, (#2), #3}
\newcommand{\zeitc}[3]{Z. Phys. {\bf C#1}, (#2), #3}
\newcommand{\mpl}[3]{Modern. Phys. Lett. {\bf A#1}, (#2), #3} 
\newcommand{\ijmp}[3]{Int. J. Modern. Phys. Lett. {\bf A#1}, (#2), #3}


\newcommand{\psibar}{\bar{\psi}}
\newcommand{\upp}{u_{+}(p,\sigma)} 
\newcommand{\vmp}{v_{-}(p,\sigma)} 
\newcommand{\umbq}{\bar{u}_{-}(q,\tau)} 
\newcommand{\vpbq}{\bar{v}_{+}(q,\tau)} 
\newcommand{\umbp}{\bar{u}_{-}(p,\sigma)} 
\newcommand{\vpbp}{\bar{v}_{+}(p,\sigma)}
\newcommand{\vmbp}{\bar{v}_{-}(p,\sigma)} 
\newcommand{\upk}{u_{+}(k,s)}
\newcommand{\vpk}{v_{+}(k,s)} 
\newcommand{\vmk}{v_{-}(k,s)} 
\newcommand{\umbk}{\bar{u}_{-}(k,s)} 
\newcommand{\upbk}{\bar{u}_{+}(k,s)}
\newcommand{\vpbk}{\bar{v}_{+}(k,s)} 
\newcommand{\upmk}{u_{\pm}(k,s)} 
\newcommand{\vpmk}{v_{\pm}(k,s)} 
\newcommand{\upmbk}{\bar{u}_{\pm}(k,s)} 
\newcommand{\vpmbk}{\bar{v}_{\pm}(k,s)} 
\newcommand{\bp}{b^\dagger_{+}(q,\tau)}
\newcommand{\bm}{b_{-}(p,\sigma)}
\newcommand{\ddp}{d^\dagger_{+}(p,\sigma)}
\newcommand{\dm}{d_{-}(q,\tau)}
\newcommand{\cbetap}{\cos{\beta(p)}}
\newcommand{\cbq}{\cos{\beta(q)}}
\newcommand{\cbk}{\cos{\beta(k)}}
\newcommand{\gfive}{\gamma_5}
\newcommand{\gfivetc}{\gamma_5 T_c}  
\newcommand{\no}{{\bf:}\Omega{\bf:}}
\newcommand{\nopq}{{\bf:}\Omega(p,q){\bf:}}
\newcommand{\vone}{{\cal V}_1}
\newcommand{\vtwo}{{\cal V}_2}
\newcommand{\gp}{G_{+}}
\newcommand{\gm}{G_{-}}
\newcommand{\ps}{\langle + \vert}
\newcommand{\ms}{\vert - \rangle }
\newcommand{\splus}{S_{+}(p)}
\newcommand{\sm}{S_{-}(p)}
\newcommand{\fnot}[1]{\!\!\not\!#1}
\newcommand{\notp}{\!\!\not\!p}
\newcommand{\calv}{{\cal V}}
\newcommand{\calh}{{\cal H}} 
\newcommand{\vmu}{V_{1\mu}(p+k)}
\newcommand{\vnu}{V_{1\nu}(p+k)}
\newcommand{\dmunu}{D_{\mu\nu}(p-k)}
\newcommand{\wplus}{\omega_{+}(p)+ \omega_{+}(k)} 
\newcommand{\wminus}{\omega_{-}(p)+ \omega_{-}(k)} 
\newcommand{\wpm}{\omega_{\pm}(p)+ \omega_{\pm}(k)} 
\newcommand{\wtwo}{2\omega_{\pm}(p)} 
\newcommand{\gb}{\bar{g}^2}
\newcommand{\svp}{\vert + \rangle }
\newcommand{\svm}{\vert-\rangle }
\newcommand{\svpm}{\vert\pm\rangle }
\newcommand{\svap}{\vert A+\rangle}
\newcommand{\svam}{\vert A-\rangle}
\newcommand{\svapm}{\vert A \pm\rangle}
\newcommand{\svpd}{\langle +\vert}
\newcommand{\svmd}{\langle -\vert}
\newcommand{\svapd}{\langle A+\vert}
\newcommand{\svamd}{\langle A-\vert}
\newcommand{\svpmd}{\langle \pm \vert}
\newcommand{\svapmd}{\langle A \pm \vert}
\newcommand{\be}{\begin{eqnarray}}
\newcommand{\ee}{\end{eqnarray}}
\newcommand{\nap}{\vert n_A + \rangle }
\newcommand{\nbp}{\vert n_B + \rangle }
\newcommand{\napd}{\langle n_A +\vert}
\newcommand{\nbpd}{\langle n_B +\vert}
\newcommand{\atp}[2]{\vert {#1}_A \bar{#2}_A + \rangle }
\newcommand{\btp}[2]{\vert {#1}_B \bar{#2}_B + \rangle }
\newcommand{\bfp}[4]{\vert {#1}_B \bar{#2}_B {#3}_B \bar{#4}_B + \rangle }
\newcommand{\atpd}[2]{\langle {#1}_A \bar{#2}_A +\vert}
\newcommand{\btpd}[2]{\langle {#1}_B \bar{#2}_B +\vert}
\newcommand{\bfpd}[4]{\langle {#1}_B \bar{#2}_B {#3}_B \bar{#4}_B +\vert}
\newcommand{\bps}[1]{b_{+}(#1)}
\newcommand{\bpsd}[1]{b^\dagger_{+}(#1)}
\newcommand{\dps}[1]{d_{+}(#1)}
\newcommand{\dpsd}[1]{d^\dagger_{+}(#1)}
\newcommand{\bms}[1]{b_{-}(#1)}
\newcommand{\bmsd}[1]{b^\dagger_{-}(#1)}
\newcommand{\dms}[1]{d_{-}(#1)}
\newcommand{\dmsd}[1]{d^\dagger_{-}(#1)}
\newcommand{\Ab}{\bar{A}}
\newcommand{\Bb}{\bar{B}}
\newcommand{\ups}[1]{u_{+}(#1)} 
\newcommand{\vps}[1]{v_{+}(#1)} 
\newcommand{\ums}[1]{u_{-}(#1)} 
\newcommand{\vms}[1]{v_{-}(#1)} 
\newcommand{\upbs}[1]{\bar{u}_{+}(#1)} 
\newcommand{\vpbs}[1]{\bar{v}_{+}(#1)} 
\newcommand{\umbs}[1]{\bar{u}_{-}(#1)} 
\newcommand{\vmbs}[1]{\bar{v}_{-}(#1)} 
\newcommand{\gone}[2]{\Gamma_1(#1,#2)}
\newcommand{\omep}[1]{\omega_+(#1)}
\newcommand{\omem}[1]{\omega_-(#1)}
\newcommand{\dep}{\Delta E _+}
\newcommand{\cbe}[1]{c_\beta(#1)}
\newcommand{\sbe}[1]{s_\beta(#1)}

\begin{titlepage}
\rightline{}
\rightline{}
\rightline{}

\vspace*{2cm}

\addtocounter{footnote}{1}

\begin{center}

{\large \bf 

Weak coupling expansion of a chiral gauge theory\\ 

\vspace{0.5cm}
on a lattice in the overlap formulation
 }
\\
\vspace{0.5cm}

\vspace{2cm}
{\sc Atsushi Yamada}\footnote{On leave of absence from: 
Department of Physics, University of Tokyo, Tokyo, 113 Japan. }  
\\
\vspace{1cm}
{\it International Center for Theoretical Physics, \\
\vspace{0.5cm}
     Miramare, Trieste, Italy  }
\vspace{2cm}

{\bf ABSTRACT}
\end{center}

Weak coupling expansion of a chiral gauge theory on a lattice 
is discussed in the overlap formulation. We analyze the fermion 
propagator and the fermion-fermion-gauge boson vertex in the one 
loop level. The chiral properties of the propagator and vertex 
are correctly preserved without tuning the parameters involved 
even after the one-loop renormalization, and the ultraviolet 
divergent parts agree with the continuum theory. Our analysis, 
together with the existing studies on the vacuum polarization 
and the gauge boson n-point functions, completes the proof of 
the renormalizability of this formulation to the one loop level. 

\end{titlepage}

\baselineskip = 0.7cm


\section{Introduction.}

\indent

The idea of the domain wall fermion \cite{kaplan} may provide a new 
possibility of regularizing a chiral fermion on a lattice 
\cite{chiral,over,nn}. 
In this approach, a Weyl fermion is simulated by a 
Dirac fermion in four plus one dimensions. 
Gauge fields $A_\mu(x)$ interacting with the 
Weyl fermion in four dimensions are 
extended to four plus one dimensions to be independent of 
the time coordinate, $A_\mu(t,x)=A_\mu(x) $ for all $t$. 
The Dirac fermion in four plus one dimensions is 
described by two different Hamiltonians ${\cal H}_+(A)$ for $t <0$ and 
${\cal H}_-(A)$ for $t>0$ where Hamiltonians $ {\cal H}_\pm (A)$ 
differ from each other only in the sign of the Dirac mass term. 
(The subscript '$\pm$' indicates the sign of the mass term.) 
Because of this step function behavior of the mass term, there arises 
one zero mode bound to the domain wall (at $t$=0), which 
behaves like a chiral fermion in four dimensions. 
This domain wall fermion inspired the overlap formulation \cite{over}. 
To move from the domain wall fermion to the overlap formulation, 
assume that the four plus one dimensional space is of Minkowskian signature, and 
consider the path integral representation of the propagator 
of the domain wall fermion in the presence of the gauge fields,  
\be
\int {\cal D} \psi {\cal D} {\bar \psi} 
\psi(t,x){\bar \psi}(t',x') \Bigr|_{t,t'=0} 
e^{i \int^{0}_{-L} {\cal L}_+(A) dt  +  i \int^{L}_{0} {\cal L}_-(A) dt },
\label{eqn:dpro} 
\ee
where ${\cal L}_\pm(A)$ are the Lagrangians corresponding to 
${\cal H}_\pm(A)$ and $L$ is the size of the fifth dimension (which is 
the time coordinate). 
The Dirac fermion $\psi(t,x)$ at $t=0$ simulates the 
Weyl fermion and thus eq. (\ref{eqn:dpro}) simulates the 
propagator for the Weyl fermion. 
Taking the limit $L \rightarrow \infty$, eq. (\ref{eqn:dpro}) will be reduced 
to, 
\be 
\svapd T\{\psi(t,x){\bar \psi}(t',x')\}\Bigr|_{t,t'=0} \svam,
\label{eqn:opro}
\ee  
where $\svapm$ are the Dirac vacua of the Hamiltonians ${\cal H}_{\pm}(A)$. 
The overlap formulation directly treats $\svapm$ and the form of the 
propagator (\ref{eqn:opro}) without invoking the four plus one dimensional 
theory. The effective action of a chiral fermion in the 
presence of the gauge fields (the determinant of the chiral Dirac operator) 
is expressed by the overlap of the two vacua,
\be
\det\{\frac{1}{2}(1 \pm \gfive)(\fnot{\partial} + ig  
{\not \!\!A} ) \}y
=e^{-\Gamma(A)} \simeq \langle A + \vert A - \rangle. 
\label{eqn:overlap}
\ee
The advantage of directly treating the vacuum overlap (\ref{eqn:overlap}) 
is that, when the gauge fields exhibit topologically non-trivial 
configurations, the overlap (\ref{eqn:overlap}) 
vanishes \cite{ins} (as the chiral determinant should), while 
if the vacuum overlap is expressed in a similar way to (\ref{eqn:dpro}) 
as a limit $L \rightarrow \infty$, it might not vanish due to possible 
contributions of the excited states of $ {\cal H}_{\pm}(A)$.   
 
Based on eq. (\ref{eqn:overlap}), treating the gauge fields $A_\mu(x)$ as 
slowly varying external fields, it has been in fact confirmed that the 
vacuum overlap $ \langle A + \vert A - \rangle$ correctly reproduces 
the chiral anomaly \cite{anomaly}, 
the vacuum polarization \cite{vac}, 
and gauge boson $n$-point functions \cite{aoki,n}. 
The validity of this formulation has been further examined recently 
including the dynamics of the gauge fields \cite{kikukawa,atsushi}. 
The fermion propagator is computed including the self-energy corrections 
due to the gauge interactions at one-loop level, 
and has been shown to remain chiral at one loop level without 
tuning the parameters involved \cite{atsushi}. 
This analysis suggests the renormalizability of this 
formulation in the presence 
of the dynamical fermions and gauge bosons. 
In this paper, we discuss the weak coupling expansion 
in the overlap formulation and present the analysis of the 
propagator and the fermion-fermion-gauge boson vertex at one loop level. 
The chiral properties of the vertex are correctly preserved as well as 
those of the 
propagator, and the ultraviolet divergent parts agree with 
the continuum theory. Our analysis, together with  
the studies \cite{vac,n,atsushi}, 
completes the proof of the renormalizability of this formulation 
at one loop level. 
The rest of the paper is organized as follows. 
In Sec. 2, we briefly review the overlap formulation. 
After we derive the Hamiltonians ${\cal H}_{\pm}(A)$ and their 
Dirac vacua $\svapm$, 
we compute the propagator (\ref{eqn:opro}) for $A_\mu(x)=0$ and analyze 
its  pole. We show that the propagator describes a chiral fermion. 
In Sec. 3, we first compute the fermion propagator including the one-loop 
self-energy corrections due to the gauge interactions, and renormalize the 
propagator. We confirm that the fermion is renormalized preserving 
the desired chiral structure without tuning the parameters involved. 
Next, we compute the vertex corrections at one-loop level, and 
show that the chiral structure and ultraviolet divergences are correctly 
reproduced. We will give some discussions in Sec. 4. 
For numerical simulations in the overlap formulation and further 
applications of this formulation, see Refs. \cite{num} and 
Refs. \cite{appl}, respectively.


\section{Formalism.}

\indent

As is discussed in the introduction, 
the Hamiltonians ${\cal H}_{\pm}(A) $ 
and their Dirac vacua $\svapm$ are the principal quantities in the overlap 
formulation. 
First we derive the Hamiltonian 
${\cal H}_{\pm}(A) $ from the four plus one dimensional Dirac Lagrangian. 
Then we express $\svapm$ as a perturbation series in the weak coupling 
expansion. 

\subsection*{The overlap Hamiltonians ${\cal H}_{\pm}(A) $
and their Dirac vacua $\svapm$.}

\indent

The Lagrangians of free Dirac fermions in four plus one dimensions are, 
\be
{\cal L}_{\pm} = \psibar(t,x)(i \sum_{A=0}^{5} \Gamma^A \partial_A \pm  
T_c  \Lambda) 
\psi(t,x), 
\label{eqn:l}
\ee
where the mass $\Lambda$ corresponds to the height of the domain wall in the 
original domain wall fermion in Ref. \cite{kaplan} and 
$T_c $ determines the fermion chirality, as will be seen later.  
The gamma matrices in four plus one dimensional Minkowskian space 
satisfy the relations, 
\be
\{\Gamma^A,   \Gamma^B \}=2 \eta_{AB},
\,\,\,\,\,
\Gamma^{0 \dagger} = \Gamma^{0}, 
\,\,\,\,\,
\Gamma^{\mu \dagger}= -\Gamma^{\mu},
\,\,\, 
(\mu =1,\cdots,4),   
\ee
with $ \eta_{AB}=(1,-1,-1,-1,-1)$, 
and they are related to the four dimensional Euclidean gamma matrices 
\cite{gam} as, 
\be
\Gamma^{0} = \gamma^{E}_5,
\,\,\,\,\, 
\Gamma^{\mu}= i\gamma^{E}_{\mu}, 
\,\,\,\,\,
\{\gamma^E_\mu,\gamma^E_\nu \}=2 \delta_{\mu\nu}, 
\,\,\,\,\,
\gamma^{E \dagger}_\mu =\gamma^E_\mu. 
\ee
(Hereafter we omit the superscript `E` of the Euclidean gamma matrices.) 
The Lagrangians (\ref{eqn:l}) lead to the Hamiltonians, 
\be
& &{\cal H}_{\pm} = \int d^4 x \psi^\dagger \gfive 
\Bigl[\sum_{\mu=1}^{4}  \gamma_\mu \partial_\mu   
\pm  T_c \Lambda   \Bigr]  \psi.
\label{eqn:Hc}
\ee
These Hamiltonians are discretized as, 
\be
{\cal H}_{\pm} &=&  \frac{1}{a}T_c(\pm \lambda + 4r  )  
\sum_{x} \psibar_n  \psi_n
- \frac{1}{2a}  \sum_{x} [ \psibar_n (T_c r-\gamma_\mu) \psi_{n+\hat{\mu}}
+  \psibar_{n+\hat{\mu}} (T_c r+\gamma_\mu) \psi_n ],
\label{eqn:hdis}
\ee
by the replacements, 
\be
x_\mu \rightarrow a n_\mu,
\,\,\,\,\, 
\psi(x) \rightarrow \psi_{n},  
\,\,\,\,\,
\int d^4x \rightarrow \sum_{x} = a^4 \sum_{n}, 
\,\,\,\,\,
\partial_\mu  \psi_n \rightarrow \frac{1}{2a} 
[ \psi_{n+\hat{\mu}} - \psi_{n-\hat{\mu}}] ,
\label{eqn:dis}
\ee
where $a$ is the lattice spacing and we 
added the Wilson term (multiplied by $T_c$),
\be 
-a^4 T_c\sum \frac{ar}{2} \psibar_n \Box \psi_n,
\,\,\,  
\Box \psi_n= \frac{1}{a^2} \sum_{\mu} \Bigl[ 
\psi_{n +\hat{\mu}} + \psi_{n -\hat{\mu}} - 2\psi_{n}  
\Bigr],
\label{eqn:wilson}
\ee
with $\Lambda = \lambda/a$. 
Here $\psibar=\psi^\dagger\gfive$ 
because of the relation $\Gamma^{0} = \gamma_5 $. 
The gauge interactions are implemented into the Hamiltonians 
(\ref{eqn:hdis}), using the link variables $U_{n,n+\hat{\mu}}$, as,  
\be
\psibar_n \psi_{n + \hat{\mu}}  \rightarrow 
\psibar_n U_{n,n+\hat{\mu}}  \psi_{n + \hat{\mu}},
\,\,\,\,\,
\psibar_{n + \hat{\mu}} \psi_{n} \rightarrow
\psibar_{n + \hat{\mu}} U_{n + \hat{\mu},n}  \psi_{n}, 
\,\,\,\,\,
U_{n + \hat{\mu},n} = U_{n,n+\hat{\mu}}^\dagger,
\label{eqn:int}
\ee
where, in the weak coupling expansion, the link valuables are expanded 
in terms of the gauge coupling $g$ with the identifications, 
\be
U_{n,n+ \hat{\mu}} = e^{iga A_\mu (n) }= 1+ iga A_\mu (n) + 
\frac{1}{2!}\{ iga A_\mu (n) \}^2 + \cdots.    
\label{eqn:link}
\ee
Here we should note that the gauge fields $ A_\mu (n)$ describe the gauge 
interactions in four dimensional Euclidean space and they are 
completely independent of the time $t$ in the four plus one dimensional sense. 
(For simplicity, we will consider the Abelian gauge theory.) 
The Hamiltonians ${\cal H}_{\pm}(A) $ are obtained by inserting 
eqs. (\ref{eqn:int}) and (\ref{eqn:link}) into eq. (\ref{eqn:hdis}). 
For the weak coupling expansion, it is convenient to use momentum 
representations.  
Performing the following Fourier transformations,  
\be
\psi_n= \int_{p} \psi(p) e^{ipan}, 
\,\,\,\,\,
\psibar_n= \int_{q} \psibar(q) e^{-iqan}, 
\,\,\,\,\,
A_\mu (n) =\int_{p} A_\mu(p) e^{ipa(n + \hat{\mu}/2)} ,
\label{eqn:fourier}
\ee
the Hamiltonians (\ref{eqn:hdis}) with 
eqs. (\ref{eqn:int}) and (\ref{eqn:link}) are, 
\be
& &{\cal H}_{\pm}(A) = \int_{p} \psi^\dagger (p) H_{\pm}(p) \psi(p) +
{\cal V}(A), 
\label{eqn:H}
\\
& &H_{\pm}(p)=\gfive\Bigl[\sum_{\mu=1}^{4} i {\tilde p}_\mu \gamma_\mu + T_c
X_{\pm}(p) \Bigr] , \,\,\,\,\,\,
X_{\pm}(p)= \pm\frac{\lambda}{a} + \frac{ar}{2} \hat{p}^2.
\label{eqn:h}
\ee
where $\tilde{p}_\mu=(1/a)\sin (p_\mu a)$, $\hat{p}_\mu=(2/a)\sin(p_\mu a/2)$ 
and the momentum integral is over the Brillouin zone $[-\pi/a,\pi/a]$. 
In eq. (\ref{eqn:H}), the first terms are the free Hamiltonians 
${\cal H}_{\pm}(A=0)$. 
The interaction term ${\cal V}$ is treated as perturbations and 
it is expanded in terms of the gauge coupling $g$ as, 
$ {\cal V}= {\cal V}_{1} + {\cal V}_{2}+ {\cal V}_{3} + \cdots$, with 
\be
& &{\cal V}_{1} = i g \int_{p,q} \psibar(p) \sum_{\mu}
V_{1\mu}(p+q) A_\mu(p-q) \psi(q),
\nonumber
\\
& &{\cal V}_2 = \frac{1}{2} a g^2 \int_{p,s,q} \psibar(p)\sum_{\mu,\nu}
V_{2\mu}(p+q)\delta_{\mu\nu}A_{\mu}(s)A_{\nu}(p-s-q)\psi(q),
\nonumber
\\
& & {\cal V}_{3} = -\frac{1}{3!}ia^2 g^3 \int_{p,s,t,q} 
\psibar(p) \sum_{\mu\nu\tau}\delta_{\mu\nu}\delta_{\mu\tau}
V_{3\mu}(p+q) A_\mu(s) A_\nu(t) A_\tau(p-s-t-q)  \psi(q),
\label{eqn:int3}
\ee
where   
$V_{1\mu}(p) = \gamma_{\mu} \cos(p_\mu a/2) -ir T_c \sin(p_\mu a/2)$, 
$V_{2\mu}(p)=  T_c r\cos(p_\mu a/2) -i\gamma_\mu \sin(p_\mu a/2) $, 
and $V_{1\mu}(p) = V_{3\mu}(p)$.

To quantize the system, we set the commutation relations,   
\be
\{\psi_{\alpha m}, \psi^\dagger_{\beta n} \} = 
\frac{1}{a^4} \delta_{\alpha\beta} \delta_{mn},
\,\,\,\,\, 
\{\psi_{\alpha m}, \psi_{\beta n} \} =
\{\psi^\dagger_{\alpha m}, \psi^\dagger_{\beta n} \} =0, 
\label{eqn:etc}
\ee
which are the equal time commutation relations for 
a four plus one dimensional Dirac fermion. 
The operator $\psi (p)$ is expanded in terms of the creation and 
annihilation operators as, 
\be
\psi(p)=  \sum_{\sigma} \Bigl[ u_{\pm}(p,\sigma) b_{\pm}(p,\sigma)
+ v_{\pm}(p,\sigma) d^\dagger_{\pm}(p,\sigma)  \Bigr],
\label{eqn:psi}
\ee    
where $u_{\pm}$ and $v_{\pm}$ are the eigenspinors of the one-particle 
free Hamiltonians $H_{\pm}(p)$ ($\sigma$ is the spin index),  
\be
& &H_{\pm}(p) u_{\pm}(p,\sigma) = \omega_\pm (p) u_{\pm}(p,\sigma),
\,\,\, 
H_{\pm}(p) v_{\pm}(p,\sigma) = -\omega_\pm (p) v_{\pm}(p,\sigma), 
\nonumber \\
& &\,\,\,\,\,\,\,\,\,\,\,\,\,\,\,\,\,\,
\omega_\pm (p) =\sqrt{{\tilde p}^2+ X^2_{\pm}(p) }, 
\label{eqn:spinor}
\ee
which satisfy the orthonormality conditions,
\be
u_{\pm}(p,\sigma)^\dagger  u_{\pm}(p,\tau)=  v_{\pm}(p,\sigma)^\dagger  v_{\pm}\
(p,\tau)= \delta_{\sigma\tau},
\,\,\,\,\,
u_{\pm}(p,\sigma)^\dagger  v_{\pm}(p,\tau)= 0.
\label{eqn:ortho}
\ee
Their explicit expressions are, 
\be
u_{\pm}(p,\sigma) =
\frac{\omega_{\pm} + X_{\pm}  -i \sum_{\mu}{\tilde  p}_\mu \gamma_\mu T_c}
{\sqrt{2\omega_{\pm}(\omega_{\pm} + X_{\pm})}} \chi(\sigma),
\nonumber \\
v_{\pm}(p,\sigma) =
\frac{\omega_{\pm} - X_{\pm}  + i \sum_{\mu}{\tilde  p}_\mu \gamma_\mu T_c}
{\sqrt{2\omega_{\pm}(\omega_{\pm} - X_{\pm})}} \chi(\sigma),
\label{eqn:uv}
\ee
where the spinor $\chi(\sigma)$ satisfies 
$\gfivetc \chi(\sigma)=\chi(\sigma)$. 
The creation annihilation operators $(b_+,d_+) $ satisfy the 
commutation relations, 
\be
& &\{b_{+}(p,\sigma),b^\dagger_{+}(q,\tau)\}= 
\{d_{+}(p,\sigma),d^\dagger_{+}(q,\tau)\}=
(2\pi)^4\delta_{\sigma\tau}\delta^4_P(p-q),
\nonumber \\
& &\{b_{+}(p,\sigma),d^\dagger_{+}(q,\tau)\}=
\{b_{+}(p,\sigma),d_{+}(q,\tau)\}=0,
\label{eqn:cm}
\ee
where $\delta_P(p-q)$ is the periodic $\delta$-function on the lattice. 
(The same relations hold also for $(b_-,d_-) $). 
The two spinor basis are related by, 
\be
& &u_{-}(p,\sigma)= \cbetap u_{+}(p,\sigma) -\sin\beta(p)
v_{+}(p,\sigma),
\nonumber \\
& &v_{-}(p,\sigma)= \sin\beta(p) u_{+}(p,\sigma)+ \cbetap
v_{+}(p,\sigma),
\label{eqn:bog1}
\ee
where $\cbetap= u^\dagger_{+}(p,\sigma) u_{-}(p,\sigma) $ is, 
\be
\cbetap =\frac{1}{\sqrt{2\omega_+ 2\omega_-}}
[\sqrt{(\omega_+ + X_+)(\omega_- +  X_-) }  +
\sqrt{(\omega_+ - X_+)(\omega_- -  X_-) }
].
\label{eqn:cb}
\ee
Eqs. (\ref{eqn:bog1}) lead to the Bogoluibov transformation 
between the two basis $(b_+,d_+)$ and $(b_-,d_-) $,  
\be
& &b_{-}(p,\sigma)= \cbetap b_{+}(p,\sigma) -\sin\beta(p)
d^\dagger_{+}(p,\sigma),
\nonumber \\
& &d^\dagger_{-}(p,\sigma)= \sin\beta(p) b_{+}(p,\sigma)+ \cbetap
d^\dagger_{+}(p,\sigma).
\label{eqn:bog}
\ee
The commutation relations between $(b_+,d_+)$ and $(b_-,d_-) $ 
are obtained from eqs. (\ref{eqn:bog}). For example,  
\be
& &\{b_{+}(p,\sigma),b^\dagger_{-}(q,\tau)\}=
\{d_{+}(p,\sigma),d^\dagger_{-}(q,\tau)\}=
(2\pi)^4 \cbetap\delta_{\sigma\tau}\delta^4_P(p-q). 
\label{eqn:cm2}
\ee
The Dirac vacua $\svpm$ for the free Hamiltonians 
$ {\cal H}_{\pm}(A=0)$ are defined as, $b_{+},d_{+}\svp =0$ and
$b_{-},d_{-}\svm =0$ and their energy eigenvalues
are denoted as $E_\pm(0)$, 
$ {\cal H}_{\pm}(A=0) \svpm = E_\pm(0) \svpm$. 
Then for the Dirac vacua $\svapm$, the eigenvalue equations, 
\be
{\cal H}_{\pm}(A) \svapm = E_{\pm}(A) \svapm, 
\label{eqn:eigena}
\ee
are solved in the form of the integral equation using the 
Dirac vacua $\svpm$ following the standard time independent 
perturbation theory, and the results are,   
\be 
\svapm &=& \alpha_\pm(A)\Bigl[1-G_\pm ({\cal V}-\Delta E_\pm)\Bigr]^{-1} \svpm,
\label{eqn:al}
\ee
where, $\Delta E_\pm = E_{\pm}(A)- E_{\pm}(0)= 
\svpmd {\cal V} \svapm / \svpmd \svapm$, 
\be
G_\pm = \sum_{n} {}' \vert n \pm \rangle \frac{1}{E_\pm(0)-E_\pm(n)} 
\langle n \pm \vert =  
\frac{1-\svpm \svpmd}{E_\pm(0)-H_\pm(0)} ,
\label{eqn:gpm}
\ee
and the sum $ \sum_{n}'$ is over all the excited states 
$\vert n \pm \rangle$ of 
${\cal H}_{\pm}(0) $ and $ E_\pm(n)$ are their energy eigenvalues. 
The normalization factors $\alpha_\pm(A)$ are determined, up to phases 
\cite{phase}, 
by the normalization conditions of 
$\svapm$ as, 
\be
\vert \alpha_\pm(A)   \vert^2= 1-
\svapmd \Bigl[{\cal V}-\Delta E_\pm\Bigr]G^2_\pm
\Bigl[{\cal V}-\Delta E_\pm\Bigr] \svapm. 
\label{eqn:alpha}
\ee

\subsection*{Fermion propagator at tree level}

\indent

Having presented $ {\cal H}_{\pm}(A)$ and $\svapm$, we now discuss the 
fermion propagator. 
In four plus one dimensions, the propagator is defined by the vacuum 
expectation value of the T-product, 
\be
T\{ \psi(t,p) \psibar(t',q)  \}=    \theta(t -t') \psi(t,p) \psibar(t',q) 
- \theta(t' -t) \psibar(t,q) \psibar(t,p).
\ee
Since $ \psi(t,p)$ at $t=0$ simulates the Weyl fermion, 
setting $t,t'=0$ and using $\theta(0)=1/2$,  
\be
T\{\psi(t,p) \psibar(t',q)\} \Bigr|_{t,t'=0} = 
\{\psi(p) \psibar(q) - \psibar(q)\psi(p)\}/2,
\ee
which we denote as $\Omega(p,q) $: 
$\Omega(p,q) =  \{\psi(p) \psibar(q) - \psibar(q)\psi(p)\}/2$. 
Then, the fermion propagator is defined by the path integral, 
\begin{eqnarray}
\frac{\int {\cal D}{\cal A}
\langle A+ \vert \Omega(p,q)\vert A- \rangle e^{-S(A)} }
{ \int {\cal D}{\cal A}    \langle A+ \vert A- \rangle e^{-S(A)} },
\label{eqn:path}
\end{eqnarray}
where $S(A)$ is
the action of the gauge field.
We find it convenient for later calculations
to decompose $\Omega(p,q)$ in the following 'normal ordered' form (using
the Bogoluibov transformation (\ref{eqn:bog})),
$\Omega(p,q) = \svpd \Omega(p,q) \svm + \nopq $,
\be
\svpd \Omega(p,q) \svm &=&  (2\pi)^4 \delta^{4}_P(p-q)
\frac{1}{2}\Bigl[ S_{+}(p) -S_{-}(p)  \Bigr],
\label{eqn:tree}
\\
\nopq&=&\frac{1}{\cbetap\cbq} \cdot
\sum_{\sigma,\tau} \Bigl[ - \upp\umbq \bp\bm
\nonumber \\
& &+  \upp \vpbq \bm\dm  + \vmp\umbq\ddp\bp
\nonumber \\
& &+ \vmp\vpbq\ddp\dm \Bigr].
\label{eqn:normal}
\ee
Eq. (\ref{eqn:tree}) is the propagator for $A=0$ and the 
normal ordered product (\ref{eqn:normal}) satisfies the relation 
$\svpd \no \svm=0 $.

Now we discuss the pole structure of the propagator (\ref{eqn:tree}) 
for the free fermions, 
\be
\frac{1}{2} \Bigl[  S_+ (p) - S_-(p) \Bigr] = 
\frac{1}{2\cbetap} \sum_{\sigma} \Bigl[ \upp\umbp - \vmp\vpbp
\Bigr].     
\label{eqn:tpro}
\ee 
The spinors $u_{\pm}, v_{\pm} $ are properly normalized vectors and can not 
yield divergences in eq. (\ref{eqn:tpro}) for any value of the momentum 
$p$. Therefore, the pole of the propagator is given by the zero of 
$\cbetap$, which occurs only  
for $\omega_+ = X_+ $ and $\omega_- = -X_- $. 
(The other case, $\omega_+ = - X_+ $ and 
$\omega_- =  X_- $, is excluded by the facts 
$\omega_+ >\omega_- >0  $ and $X_+ > X_- $.) 
The necessary and sufficient condition for 
the pole of the propagator is $\tilde{p}_\mu=0  $ and $X_- <0  $. 
At the origin of the Brillouin zone $p \simeq 0$, 
$\tilde{p}_\mu=0$ and $X_- =-\lambda/a$, 
which is the pole of the propagator. 
At each corner of the Brillouin zone $p_\mu = \pm \pi/a + q_\mu$, 
$\tilde{p}_\mu = 0$ and $X_-= (-\lambda +2rn)/a$, 
where $n=1,\cdots,4$ 
is the number of momentum components which lie near the corner of the
Brillouin zone. Thus, for the range of the parameters 
$\lambda < 2r$, each corner does not leads to the pole of the propagator.    
In fact, for $p \simeq 0$, 
$\omega_{\pm}(p) = {\lambda}/{a} + {\cal O}(a)$, 
$\cbetap = {a \sqrt{p^2}}/{\lambda} +{\cal O}(a^3)$, 
\be
u_{+}(p,\sigma) = 
[1-\frac{a}{2\lambda} i\notp T_c + {\cal O}(a^2)]\chi(\sigma), 
\,\,\,\,\,
u_{-}(p,\sigma)= 
\frac{1}{\sqrt{p^2}}[- i\notp T_c  + \frac{a}{2\lambda}p^2 +{\cal O}(a^2)]
\chi(\sigma),
\nonumber \\
v_{-}(p,\sigma))= 
[1+\frac{a}{2\lambda} i\notp T_c + {\cal O}(a^2)]\chi(\sigma), 
\,\,\,\,\,
v_{+}(p,\sigma) = 
\frac{1}{\sqrt{p^2}}[ i\notp T_c  + \frac{a}{2\lambda}p^2+{\cal O}(a^2)]
\chi(\sigma),
\label{eqn:uvl}
\ee
which lead to, 
\begin{eqnarray}
S_{\pm}(p) \simeq \frac{\lambda}{a}\frac{1}{p^2}
\Bigr[ \frac{1}{2}(1+\gfivetc)(\mp i \notp) +\frac{a}{2\lambda}p^2 \gfive
\Bigl] ,
\label{eqn:propagator}
\end{eqnarray}
and thus the propagator (\ref{eqn:tpro}) 
describes a chiral fermion
in this region.
At each corner of the Brillouin zone, $p_\mu \simeq  \pm \pi/a + q_\mu$,
\be
\cbetap= 1 + \frac{1}{\sqrt{(4 n^2 r^2 - \lambda^2)}}{\cal O}(a^2q^2),   
\ee
and the propagator takes the following form,
\begin{eqnarray}
\frac{1}{2}\Bigl[ S_{+}(p) -S_{-}(p)  \Bigr]
\simeq \frac{1}{ \sqrt{(4 n^2 r^2 - \lambda^2)} + {\cal O}(a^2q^2) }
(c_1+c_2\gfive),
\label{eqn:propagatorp}
\end{eqnarray}
where $c_{1,2}$ are constants. Thus, 
the chiral non-invariant
contributions coming from each corner are suppressed
due to the $\sqrt{ (4 n^2 r^2-\lambda^2)}$ mass.


\section{The weak coupling expansion.}

\indent 

Next consider the effects of the gauge interactions at one loop level 
in the weak coupling expansion. 

\subsection*{Fermion propagator}

\indent 

First consider the fermion propagator. 
To obtain fermion propagator at one loop level, the expression 
(\ref{eqn:path}) 
should be expanded up to the order $g^2$. 
Inserting the decomposition of $\Omega(p,q)$ into eq. (\ref{eqn:path}), 
eq. (\ref{eqn:tree}) yields the 
propagator at tree level and all the 
quantum corrections arise from $\langle A+ \vert \no \vert A- \rangle$. 
Expanding $\svapm$ in the perturbation series,  
\be
\svapm = \alpha_\pm(A) \Bigl[ \svpm + G_\pm {\cal V} \svpm + 
G_\pm \Bigl\{   {\cal V} -\Delta E_\pm \Bigr\} G_\pm {\cal V} 
\svpm + \cdots \Bigr], 
\ee 
the quantum correction 
$\langle A+ \vert \no \vert A- \rangle$,
up to the order $g^2$, is, 
\begin{eqnarray}
& &\langle A+ \vert \no \vert A- \rangle =  
\ps \no \gm \vtwo \ms +  \ps \vtwo \gp \no \ms + 
\ps \vone \gp \vone \gp \no \ms
\nonumber \\
& &+
\ps \vone \gp \no \gm \vone \ms 
+  
\ps \no \gm \vone \gm \vone \ms .  
\label{eqn:tot}
\end{eqnarray}
(The terms containing $\Delta E_\pm$ are neglected since they are 
of order ${\cal O}(g^2) $.)  
These terms are evaluated by rewriting the fermion operators in 
$\vone$ and $\vtwo$ in terms of 
the creation and annihilation operators defined in eq. (\ref{eqn:psi}) 
and using the commutation relations. 
As an example, we evaluate the third term in eq (\ref{eqn:tot}). 
Using eq. (\ref{eqn:gpm}), 
\be
\ps \vone \gp \vone \gp \no \ms
&=&
\sum_{n_A,n_B}{}' \ps \vone \nap \frac{1}{E_{+}(0)-E_{+}(n_A)} 
\napd \vone \nbp \nonumber \\ 
& &\cdot \frac{1}{E_{+}(0)-E_{+}(n_B)} \nbpd\no \ms. 
\label{eqn:ex1}
\ee
Here, $\vert n_A +\rangle$ should be two particle states, 
\be
 \nap = \atp{1}{1}    = \bpsd{1_A}\dpsd{\bar{1}_A} \svp,
\ee
where we use the abbreviation 
$ \bpsd {1_A}=  \bpsd{{p_1}_A,{\sigma_1}_A}$ e.t.c. 
For $\vert n_B +\rangle $, both two and four particle states contribute. 
We only consider four particle states in our example,  
\be
\nbp = \bfp{1}{1}{2}{2} =  
\bpsd{2_B}\dpsd{\bar{2}_B} \bpsd{1_B}\dpsd{\bar{1}_B} \svp. 
\label{eqn:nb}
\ee 
The matrix elements of $\vone$ between relevant states are given by, 
\be
\svpd \vone \atp{1}{1} &=& \vpbs{\bar{1}_A} \gone{\bar{1}_A}{1_A}
\upbs{1_A},
\nonumber\\
\atpd{1}{1} \vone \ \bfp{1}{1}{2}{2} 
&=& 
\svpd \vone \btp{2}{2} \atpd{1}{1}   1_B \bar{1}_B + \rangle 
- (1_B \leftrightarrow 2_B )
- (\bar{1}_B \leftrightarrow \bar{2}_B )
\nonumber \\
& &
+ ( 1_B   \bar{1}_B \leftrightarrow 2_B \bar{2}_B  ),
\label{eqn:matrix}
\ee
where, 
\be
& &\gone{p}{q}=
i g \sum_{\mu}
V_{1\mu}(p+q) A_\mu(p-q),
\,\,\,\,\,
\atpd{1}{1}  1_B \bar{1}_B + \rangle = 
\delta(1_A,1_B) \delta(\bar{1}_A,\bar{1}_B),
\nonumber
\\
& &\delta(1_A,1_B)= (2\pi)^4 \delta_{\sigma_{1{_A}} \sigma_{1{_B}}} \delta^4_P
(p_{1{_A}}- p_{1{_B}}). \label{eqn:g}
\ee
Inserting eqs. (\ref{eqn:matrix}) into eq. (\ref{eqn:ex1}),
\be
\ps \vone \gp \vone \gp 
&=& \frac{1}{ 2! 2! }   
\int_{1_B\bar{1}_B 2_B \bar{2}_B } 
\Bigl[ 
\Bigl\{
\frac{-1}{\omep{1_B}+\omep{\bar{1}_B}} \Bigr\} \svpd \vone \btp{1}{1}
\nonumber \\ 
& &
\cdot
\Bigl\{ 
\frac{-1}{ \omep{1_B}+\omep{\bar{1}_B}+\omep{2_B}+\omep{\bar{2}_B}}
\Bigr\}
\svpd \vone \btp{2}{2}
\nonumber \\ 
& &
- (1_B \leftrightarrow 2_B )
- (\bar{1}_B \leftrightarrow \bar{2}_B )
+ ( 1_B   \bar{1}_B \leftrightarrow 2_B \bar{2}_B  )
\Bigr]  
\nonumber \\
& &\cdot \bfpd{1}{1}{2}{2}.
\label{eqn:pint}
\ee 
The normal ordered part is evaluated using eq. (\ref{eqn:normal})
as,
\be
\bfpd{1}{1}{2}{2} \no \svm 
&=& - \sum_{\sigma,\tau}\frac{1}{\cbe{p} \cbe{q}}
\vmp \umbq \delta(2_B,q\tau) \delta(\bar{2}_B,p\sigma)
\btpd{1}{1} - \rangle 
\nonumber \\
& &+ (1_B \leftrightarrow 2_B )
+ (\bar{1}_B \leftrightarrow \bar{2}_B )
- ( 1_B   \bar{1}_B \leftrightarrow 2_B \bar{2}_B  ), 
\label{eqn:pnormal}
\\
& &\btpd{1}{1} - \rangle
=\frac{\sbe{1_B}}{\cbe{1_B}} \delta(1_B,\bar{1}_B).
\nonumber 
\ee
Combining eqs. (\ref{eqn:pint}) and (\ref{eqn:pnormal}), 
\be
& &\ps \vone \gp \vone \gp \no \ms 
= 
\frac{2!2!}{2!2!} \int_{1_B\bar{1}_B 2_B \bar{2}_B} 
\Bigl\{
\frac{1}{\omep{1_B}+\omep{\bar{2}_B}} 
\Bigr\}\svpd \vone \btp{1}{2}
\nonumber \\
& &\cdot \Bigl\{
\frac{1}{ \omep{1_B}+\omep{\bar{2}_B}+\omep{2_B}+\omep{\bar{1}_B}}
\Bigr\}
\svpd \vone \btp{2}{1}
\frac{1}{\cbe{p}\cbe{q}}
\vmp \umbq 
\nonumber \\
& &  \cdot
\delta(2_B,q\tau) \delta(\bar{2}_B,p\sigma)
\btpd{1}{1} - \rangle  + (1\,\,\,more\,\,\, term),
\nonumber \\
& &= S_{-}(p) \int_{p_{1_B}}
\Bigl\{\frac{1}{\omep{1_B}+\omep{p}}\Bigr\}\Bigl\{
\frac{1}{ \omep{p}+\omep{q}+2\omep{1_B}}\Bigr\}\gone{p}{1_B}
\nonumber \\
& &\cdot  T_{+}(1_B) \gone{1_B}{q} S_{+}(q) 
+ (1\,\,\,more \,\,\,term),
\label{eqn:self}
\ee
where,  
\be
T_{+}(p)= \sum_{\sigma} \frac{\sbe{p}}{\cbe{p}} 
\upp \vpbp,  
\ee
and we only considered the one-particle irreducible contributions. 
(The one-particle reducible contributions 
vanish because there is no tadpole contribution for the gauge fields. 
Namely, the terms containing the combination, for example, 
$\svpd \vone \btp{1}{1} \btpd{1}{1} - \rangle$, vanish after the momentum 
integration because of the discrete lattice symmetry.) 

Now, $\Gamma_1(p,q)$ is rewritten in terms of the gauge fields by 
eq. (\ref{eqn:g}), and the path integral over the gauge fields at 
the stage of eq. (\ref{eqn:path}) leads to the following propagator, 
\be
\langle A_\mu (p) A_\nu (q) \rangle
= (2\pi)^4 \delta^4_P (p+q) e^{-i(p+q)_\nu a/2   } D_{\mu\nu}(p), 
\,\,\,\,\,
D_{\mu\nu}(p)=  \frac{1}{\hat{p}^2}
(\delta_{\mu\nu} - (1-\xi)\frac{\hat{p}_\mu\hat{p}_\nu }{\hat{p}^2}),
\label{eqn:gaugepro}
\ee
for the bilinear of the gauge fields $A_\mu(p) A_\nu(q)$. 
Summing up all the eight terms, 
the contributions of the fourth term in eq. (\ref{eqn:tot})
are expressed as  
$- (2\pi)^4 \delta(p-q)S_{-}(p) a \Sigma(p)S_{+}(p)$, 
where 
\begin{eqnarray}
\Sigma(p) &=& \frac{1}{a} \gb \sum_{\mu,\nu} \int_{k} 
\Bigl[ \frac{1}{\wpm} \Bigr]^2
\vmu\ T_{+}(k)  \vnu \dmunu.
\label{eqn:sigma}
\end{eqnarray}
Here we explicitly factor out the lattice spacing $a$ so that
$\Sigma(p) $ has the correct dimension (one) of the self-energy for
fermions. As is seen from above, this term is described by the 
Feynman diagram Fig. 1. (a).

Next we compute eq. (\ref{eqn:sigma}) in the limit $a \rightarrow 0$. 
We will follow the strategy developed in Ref. \cite{kawai,kawamoto,smit}. 
First to see the divergences associated with the limits 
$a \rightarrow 0$, we factor out the $a$ dependence
from the loop momentum integration by a rescaling of the loop momentum
$k \rightarrow \bar{k}=ka$. Then the limit $a \rightarrow 0$ produces 
the linear and logarithmic divergences. 
Here, the logarithmic divergence is associated with the infrared 
divergence of the $\bar{k}$ integration in the limit $ap \rightarrow 0$. 
Such an infrared divergence takes place only in the region $\bar{k} \simeq 0$ 
because of the pole structure of the propagators (\ref{eqn:propagator}) 
and (\ref{eqn:gaugepro}). 
To evaluate these divergent terms, 
we expand the integrand with respect to the external momentum $p$ around 
$p \simeq 0$. Then, the zero and 
first terms of the Taylor expansion contain 
linear and logarithmic divergences, respectively. 
As a consequence of the expansion around $p \simeq 0$, 
the $\bar{k}$ integration also develops infrared divergences associated with 
$p \rightarrow 0$ (not $a \rightarrow 0$), so 
we regularize infrared divergences by dimensional regularization from the 
beginning. (Another way of evaluating eq. (\ref{eqn:sigma}) 
in the limit $a \rightarrow 0$ is to divide the integration region into 
two pieces, which is described in Ref. \cite{smit}.)

Starting from the expression (\ref{eqn:sigma}), first we analytically 
continue the space-time dimension $D$ {\it larger than } 4 
to regularize infrared divergences and introduce 
an arbitrary parameter $\kappa$ of mass dimension one to 
keep the dimensions of the vertices unchanged \cite{dimensional}, 
\be
\int^{\pi/a}_{-\pi/a} \Bigl(\frac{dk}{2\pi}\Bigr)^4
\rightarrow 
\kappa^{2\varepsilon} \int^{\pi/a}_{-\pi/a} \Bigl(\frac{dk}{2\pi}\Bigr)^D,
\,\,\,\,\, \varepsilon=\frac{4-D}{2},
\label{eqn:dim}
\ee
Then rescaling the loop momentum $k \rightarrow \bar{k}=ka$, 
\be 
\Sigma(p) &=& \frac{1}{a} g^2 \sum_{\mu,\nu} (\kappa a)^{2\varepsilon}
\int^{\pi}_{-\pi} \Bigl(\frac{d\bar{k}}{2\pi}\Bigr)^D 
\Bigl[ \frac{1}{ \tilde{\omega}_{+}(\bar{k}) +\tilde{\omega}_{+}(ap) }\Bigr]^2
\tilde{V}_{1\mu}(ap+\bar{k}) \tilde{T}_+(\bar{k}) 
\nonumber \\
& &\cdot \tilde{V}_{1\nu}(\bar{k}+ap) \tilde{D}_{\mu\nu}(ap-\bar{k}),
\label{eqn:rsigma}
\ee 
where the rescaling of each quantity is, 
\be
\omega_{+}(k)=  \frac{1}{a}\tilde{\omega}_{+}(\bar{k}),
\,\,\,\,\,
V_{1\mu}(k)=\tilde{V}_{1\mu}(\bar{k}), 
\,\,\,\,\,
T_{+}(k)=\tilde{T}_+(\bar{k}),
\,\,\,\,\,
D_{\mu\nu}(k) = {a^2}\tilde{D}_{\mu\nu}(\bar{k}).  
\label{eqn:res}
\ee    
Next, expanding eq. (\ref{eqn:rsigma}) with respect to the external momentum 
$p$, the momentum integral is finite for the zero term, 
and gives rise to, 
\be
\Sigma(0)= \frac{1}{a} \Bigl[
\sigma_{1}(\lambda,r,\xi) + \sigma_{2}(\lambda,r,\xi)
\gfive \Bigr],
\label{eqn:limita}
\ee
where $ \sigma_{1,2}(\lambda,r,\xi)$ are the constants. 
For the first term, 
\be
 \sum_{\mu} p_\mu 
\frac{\partial \Sigma(p)}{\partial p_\mu} \Bigl|_{p =0}, 
\ee
only the derivative of $\tilde{D}_{\mu\nu}(ap-\bar{k})$ leads to the 
integral which gives rise to the logarithmic (infrared) divergence, 
and the derivative of the other terms yields, after the integration, 
finite terms in the limit $\varepsilon$, $a$ $\rightarrow 0$. 
Computing the integral, we obtain, 
\be
\sum_{\mu} p_\mu 
\frac{\partial \Sigma(p)}{\partial p_\mu} \Bigl|_{p =0} 
&\rightarrow& 
\frac{1}{\lambda}\frac{1}{4} 
\frac{\gb}{16\pi^2} \{1-(1-\xi)\} 
\{ \frac{1}{\varepsilon} + \log(\kappa^2a^2) \} 
\nonumber \\
& &\cdot \frac{1}{2}(1-\gfivetc)i\notp + finite\,\,\, term. 
\label{eqn:limitb}
\ee
The remaining part, 
\be
\Sigma(p) - \Sigma(0) - \sum_{\mu} p_\mu 
\frac{\partial \Sigma(p)}{\partial p_\mu} \Bigr|_{p =0}, 
\label{eqn:finite}
\ee
is ultraviolet finite for $a \rightarrow 0$, but is infrared divergent 
for $\varepsilon \rightarrow0 $. 
In fact, because of the subtraction of the first two terms of the 
Taylor expansion, eq. (\ref{eqn:finite}) would have been 
of order $a$ if there were no infrared divergences at 
$ap \rightarrow 0$. 
Dividing the integration region into parts where the integral 
is regular in the limit $ap \rightarrow 0$ and 
parts where the integral exhibits an infrared divergence, 
the contribution of the former is irrelevant, 
while that of the latter is finite for $a \rightarrow 0$. 
Because of the structure of the 
poles of the propagators (\ref{eqn:propagator}) and (\ref{eqn:gaugepro}) 
such an infrared divergence can occur only at $\bar{k} \simeq 0$ 
in the limit $ap \rightarrow 0$, and the contribution of this region 
is evaluated by expanding the integrand around $\bar{k} \simeq 0 $ (and 
$ap \simeq 0$). The result of the calculation is,
\be
& &\Sigma(p) - \Sigma(0) - \sum_{\mu} p_\mu 
\frac{\partial \Sigma(p)}{\partial p_\mu} \Bigr|_{p =0} 
\nonumber \\  
\rightarrow 
& &-
\frac{1}{\lambda}\frac{1}{4} \frac{g^2}{16\pi^2} \{1-(1-\xi)\}  
\{ \frac{1}{\varepsilon}   + \log(\kappa^2/p^2) \}\frac{1}{2}
(1-\gfivetc)i\notp.
\label{eqn:limitc}
\ee  
Summing up eqs. (\ref{eqn:limita}), (\ref{eqn:limitb}) 
and (\ref{eqn:limitc}), we obtain, 
\be
\Sigma(p)&=&  \frac{1}{a} \Bigl[
\sigma_{1}(\lambda,r,\xi) + \sigma_{2}(\lambda,r,\xi)
\gfive \Bigr] +\frac{1}{\lambda}\frac{1}{4} \frac{g^2}{16\pi^2} 
\{1-(1-\xi)\}\log(p^2 a^2)
\nonumber \\
& & \cdot
\frac{1}{2}
(1-\gfivetc)i\notp + (finite \,\,\,terms).
\label{eqn:sigmatot}
\ee 
The infrared divergences at $\varepsilon \rightarrow 0$ are cancelled in eq. 
(\ref{eqn:sigmatot}). 
The contribution to the tree level propagator is given by, 
\be 
& &-S_{-}(p) a \Sigma(p)S_{+}(p)= 
-\frac{\lambda}{a}\frac{1}{p^2}
\Bigl[ \frac{1}{2}(1+\gfivetc)(+ i \notp) 
+\frac{a}{2\lambda}p^2 \gfive \Bigr] a 
\nonumber \\
& &
\cdot \Bigl[
\frac{1}{a} \Bigl\{     
\sigma_{1} + \sigma_{2} \gfive \Bigr\}   + 
\frac{1}{\lambda}\frac{1}{4} \frac{g^2}{16\pi^2} 
\{1-(1-\xi)\}\log(p^2 a^2)\frac{1}{2}
(1-\gfivetc)i\notp + (finite \,\,\,terms) \Bigr] 
\nonumber \\
& &
\cdot 
\frac{\lambda}{a}\frac{1}{p^2}\Bigl[\frac{1}{2}(1+\gfivetc)(-i \notp) + 
 \frac{a}{2\lambda}p^2 \gfive \Bigr] 
\nonumber \\ 
\rightarrow & &
\Bigl[ \sigma_1 T_c  +   \frac{1}{\lambda}\frac{1}{4} \frac{g^2}{16\pi^2}
\{1-(1-\xi)\}\log(p^2 a^2) + finite\,\,\, terms + {\cal O}(a) \Bigr]
\nonumber \\
& &\cdot 
\frac{\lambda}{a}\frac{1}{2}(1+\gfivetc)\frac{-i \notp}{p^2}. 
\label{eqn:wavefini}
\ee
From this expression, we can see that only an ultraviolet divergence of 
$\Sigma(p) $ more severe than quadratic can invalidate the 
chiral property of the regularized fermion, and  
the logarithmically divergent term in eq. (\ref{eqn:sigmatot}) directly 
appears in the wave function renormalization factor as 
a logarithmic divergence. 

If we take two particle contributions for 
$\vert n_B +\rangle $ in eq. (\ref{eqn:nb}), 
we obtain, instead of eq. (\ref{eqn:sigma}), 
\be
\Sigma(p) &=& - \frac{1}{a} g^2 \frac{1}{2 \omep{p}}
\sum_{\mu,\nu}
\int_{k}
\frac{1}{ \omep{p} + \omep{k} }
\vmu
\nonumber
\\
& &\cdot \sum_{s} \Bigl[ \upk\upbk - \vpk \vpbk \Bigr] \vnu \dmunu.
\label{eqn:pib}
\ee
This term does not contain logarithmic divergences since there appears no 
infrared divergence for the first term of the Taylor expansion after the 
rescaling and expansion. 

The other terms in eq. (\ref{eqn:tot}) are evaluated in the same way. 
(The results are given in Ref. \cite{atsushi}). 
The first two terms are expressed by the Feynman diagram Fig. 1. (b), and 
do not lead to the logarithmic divergence at the level of eq. 
(\ref{eqn:wavefini}), while the fourth and fifth terms 
correspond to the Feynman diagram Fig. 1. (a), and yield the logarithmic 
divergences. 
Summing up all the contributions as well as the propagator at tree level, 
the propagator at one loop level is, 
\be
\frac{\lambda}{a}\frac{1}{2}(1+\gfivetc)\frac{-i\notp}{p^2} 
\Bigl[ 1 + \frac{\gb}{16\pi^2}\{1-(1-\xi)\}
(\frac{1}{4}+ \frac{1}{2}+  \frac{1}{4}    )( \log a^2p^2 + finite
\,\,\,terms)\Bigr],  
\label{eqn:pone}
\ee
where the contributions of the third and fourth and fifth terms to the 
logarithmically divergent term are $1/4$, $1/2$ and $1/4$, respectively. 
From this expression, we see that the chirality of the regularized fermion 
is properly preserved and the wave function renormalization factor is, 
\be
Z_2 = 1+ \frac{g^2}{16\pi^2}\Bigl[1-(1-\xi) \Bigr] 
( \log a^2\mu^2 + const),
\ee
where $\mu$ is the renormalization scale. 
The divergent part of the wave function renormalization factor agrees 
with that of the continuum theory.

\subsection*{Vertex correction}

\indent

Next we consider fermion-fermion-gauge boson vertex. 
The three point function is given by, 
\be
\frac{\int {\cal D}{\cal A}
\langle A+ \vert \Omega(p,q)\vert A- \rangle A_\mu(t)e^{-S(A)} }
{ \int {\cal D}{\cal A}    \langle A+ \vert A- \rangle e^{-S(A)} }.
\label{eqn:ver}
\ee
To obtain the vertex up to the one loop level, eq. (\ref{eqn:ver}) 
should be expanded up to the order $g^3$. 
The procedure of the calculation is same to the case of the propagator. 
Inserting the decomposition of $\Omega(p,q)$ 
into eq. (\ref{eqn:ver}), the first part $\svpd \Omega(p,q) \svm$ 
vanishes after the path integration over the gauge fields. 
The normal ordered part $\svapd \nopq \svam$ is evaluated up to 
the order $g^3$ by expanding $\svapm$ with respect to 
the gauge coupling $g$ in eq. (\ref{eqn:al}). 
Up to this order, 
\be
& &\svapd \no \svam =  \alpha^*_+(A) \alpha_-(A) 
\Bigl\{ 
\svpd {\cal V} G_+ \no \svm + \svpd \no G_- {\cal V} \svm 
+\svpd {\cal V}  G_+ {\cal V} G_+ \no \svm 
\nonumber \\
& &
+ \svpd {\cal V} G_+ \no G_- {\cal V} \svm 
+ \svpd \no G_- {\cal V} G_- {\cal V} \svm 
+ \svpd {\cal V}  G_+ {\cal V} G_+ {\cal V} G_+ \no \svm 
\nonumber \\
& &
+ \svpd {\cal V}  G_+ {\cal V} G_+ \no G_- {\cal V} \svm 
+ \svpd {\cal V}  G_+ \no G_- {\cal V} G_- {\cal V} \svm 
+ \svpd \no G_- {\cal V} G_- {\cal V} G_- {\cal V} \svm 
\nonumber\\ 
& & 
- \Delta E_+ \svpd {\cal V} G^2_+ \no \svm 
- \Delta E_- \svpd \no G^2_- {\cal V} \svm 
\Bigr\}.
\label{eqn:totv}
\ee
The first two terms $\svpd {\cal V} G_+ \no \svm$ and 
$ \svpd \no G_- {\cal V} \svm$ with ${\cal V}=\vone$ lead to the 
vertex at tree level after the path integration over $A_\mu$, 
\be
& &(2\pi)^4 \delta_P (p+t-q) 
\sum_{\rho}
D_{\mu\rho}(t) e^{-\frac{i}{2}(p+t-q)_\rho a }
\Bigl\{ 
S_-(p) \frac{a}{\lambda} \Gamma_\rho(p,q) S_+(q) 
+
S_+(p) \frac{a}{\lambda} \Gamma_\rho(p,q) S_-(q) 
\Bigr\}, 
\nonumber \\
& &\,\,\,\,\,\,\,\,\,\,\,\, \Gamma_\rho(p,q) \rightarrow 
\frac{1}{2}ig \frac{1}{2}(1-\gfive T_c)\gamma_\rho,
\,\,\,\,\,(a \rightarrow 0).
\label{eqn:vertree}
\ee    
For the radiative corrections we only need to evaluate one-particle 
irreducible contributions. 
The vertex corrections exhibit logarithmic divergences, which associate 
with infrared divergences of the integral after the rescaling of the 
loop momentum in our strategy. 
Therefore, in this case, we essentially look for terms which 
have enough poles, or equivalently $\cbetap^{-1}$ factors, to exhibit 
infrared divergences. 
Such terms are obtained only from the interaction ${\cal V}_1$, 
and the contributions of ${\cal V}_2$ and ${\cal V}_3$ yield only 
finite contributions. 

For example, we evaluate a term containing ${\cal V}_2$, 
$\svpd  G_+ \vone  G_+ \vtwo \no \svm$. This term leads to the expression, 
\be
& &\svpd  G_+ \vone  G_+ \vtwo \no \svm \rightarrow 
\frac{2!2!}{2!2!} \int_{2_B} 
\Bigl\{ \frac{1}{\omep{q}+\omep{2_B}}\Bigr\}  
\Bigl\{ \frac{1}{\omep{q}+\omep{p} +2\omep{2_B}}  \Bigr\} 
\nonumber \\
& &\cdot S_-(p) \Gamma_2(p,2_B) T_+(2_B)\Gamma_1(2_B,q)S_+(q)
+(1\,\,\,more\,\,\,term), 
\label{eqn:convtwo}
\ee
where, 
\be 
& &\Gamma_1(p,q) = i g \sum_{\mu}
V_{1\mu}(p+q) A_\mu(p-q),
\\
& &\Gamma_2(p,q) = \frac{1}{2} a g^2 \int_{s} \sum_{\mu,\nu}
V_{2\mu}(p+q)\delta_{\mu\nu}A_{\mu}(s)A_{\nu}(p-s-q),
\ee
and we have considered only the four particle contributions for 
the evaluation of the second $G_+$ between $\vone$ and $\vtwo$. 
(The two particle contributions for second $G_+$ do not give rise to 
enough pole factors to exhibit infrared divergences.)  
After the path integration over the gauge fields, the first term in eq. 
(\ref{eqn:convtwo}) yields the following one-particle irreducible 
contribution, 
\be 
& & (2\pi)^4 \delta_P (p+t-q) 
\sum_{\rho}
D_{\mu\rho}(t) e^{ -\frac{i}{2}(p+t-q)_\rho a }
\Bigl\{
S_-(p) \frac{a}{\lambda}  \Gamma_\rho(p,q) S_+(q) 
\Bigr\} 
,
\nonumber \\ 
& & \Gamma_{\rho}(p,q) =  
\frac{\lambda}{a} ig \frac{1}{2} a g^2 
\int_{k} 
\Bigl\{ \frac{1}{ \omep{p}+\omep{k}}           \Bigr\}  
\Bigl\{ \frac{1}{ \omep{p}+\omep{q}+2\omep{k}} \Bigr\} 
\nonumber \\ 
& &\cdot \sum_{\nu} V_{1 \nu}(p+k) T_+(k) V_{2\rho}(k+q)D_{\nu\rho}(p-k). 
\ee 
This contribution is 
described by the Feynman diagram in Figs. 2. (a). 
Rescaling the momentum variable $k \rightarrow {\bar k}=ak$, 
we only need to evaluate the zero term of the Taylor expansion with respect 
to the external momenta $p$ and $q$ to compute possible divergent terms 
in the limit $a \rightarrow 0$, which is, 
\be 
\Gamma_\rho(p=0,q=0)= \frac{\lambda}{a}ig \frac{1}{4}a g^2 
\int_{{\bar k}} 
\Bigl\{\frac{1}{\lambda+\tilde{\omega}_+({\bar k})}\Bigr\}^2 
\sum_{\nu} \tilde{V}_{1 \nu}({\bar k}) \tilde{T}_+({\bar k}) 
\tilde{V}_{2\rho}({\bar k})\tilde{D}_{\nu\rho}(-{\bar k}).
\ee
The integral over ${\bar k}$ is (infrared) finite because of the 
structure of the interaction vertices, 
and thus this term only yields the finite contributions.  
The contributions of the terms $\svpd {\cal V}_3 G_+ \no \svm$ and
$ \svpd \no G_- {\cal V}_3 \svm$ are described by the Feynman diagram 
Fig. 2. (b) and yield only finite corrections to the vertex. 
In this way, the interactions ${\cal V}_2$ and ${\cal V}_3$ lead 
only to the finite terms (contact terms). 

The logarithmically divergent terms are obtained from the following terms in 
eq. (\ref{eqn:totv}),   
\be 
& &\svpd {\cal V} G_+ {\cal V}  G_+ {\cal V} G_+  \no \svm
+ 
\svpd {\cal V} G_+ {\cal V}  G_+ \no G_- {\cal V} \svm
+
\svpd {\cal V} G_+ \no G_- {\cal V} G_- {\cal V} \svm
\nonumber \\
& &+\svpd \no G_- {\cal V} G_- {\cal V} G_- {\cal V} \svm,
\,\,\,\,\,\,\,\,\,\, ({\cal V}= {\cal V}_1). 
\label{eqn:logv}
\ee
We evaluate, as an example, 
$\svpd {\cal V}_1  G_+ {\cal V}_1  G_+ {\cal V}_1 G_+ \no \svm$. 
Using the technique described in the previous subsection, 
\be 
& &\svpd {\cal V}_1  G_+ {\cal V}_1  G_+ {\cal V}_1 G_+ \no \svm 
\rightarrow \frac{3!3!}{3!3!}\Bigl[ \int_{1_C 2_C} 
\Bigl\{ \frac{1}{\omep{1_C}+\omep{2_C}} \Bigr\}
\Bigl\{ \frac{1}{\omep{1_C}+\omep{q} +2\omep{2_C}} \Bigr\}
\nonumber \\
& &
\cdot \Bigl\{ \frac{1}{2\omep{1_C}+2\omep{2_C}+\omep{p}+\omep{q}} \Bigr\}
S_-(p) \Gamma_1(p,1_C) T_+(1_C) \Gamma_1(1_C,2_C) T_+(2_C) 
\nonumber \\
& &
\cdot \Gamma_1(2_C,q) S_+(q) + (5 \,\,\,terms) \Bigr],  
\label{eqn:vlog}
\ee
where we have taken two, four and six particle states 
in the evaluation of the first, second and third $G_+$ from the left, 
respectively. (The contributions of the other states do not lead to the 
divergent terms. This situation is quite similar to the case of 
eq. (\ref{eqn:pib}), which does not exhibit the infrared divergence for the 
first term of the Taylor expansion. These terms do not have enough 
$\cbetap^{-1} $ factors to develop logarithmic divergences. ) 
In eq. (\ref{eqn:vlog}) the six terms are different with each other only in 
the momentum arguments in $\omega_+$. For example, there is a term 
proportional to, 
\be
& &\Bigl\{
\frac{1}{\omep{q}+\omep{1_c}} 
\Bigr\}
\Bigl\{
\frac{1}{\omep{q}+\omep{1_c} + \omep{p}+\omep{2_c}}  
\Bigr\}
\nonumber \\
& &\cdot \Bigl\{
\frac{1}{\omep{q}+2\omep{1_c} + \omep{p}+ 2\omep{2_c}} 
\Bigr\}.
\ee
Such differences are irrelevant for the calculation of the divergent 
parts of the vertex function, as will be seen later. 
Multiplying eq. (\ref{eqn:vlog}) with $A_\mu(t)$, and performing the 
path integral over gauge fields, we obtain (for the one-particle irreducible 
contributions) 
\be 
& & (2\pi)^4 \delta_P (p+t-q) 
\sum_{\rho} D_{\mu\rho}(t) e^{-\frac{i}{2}(p+t-q)_\rho a }
\Bigl\{
S_-(p) \frac{a}{\lambda} \Gamma_\rho(p,q) S_+(q)
\Bigr\},
\nonumber \\
& & \Gamma_\rho(p,q)= \frac{\lambda}{a} (ig)^3 \int_{k} 
\Bigl\{
\frac{1}{\omep{q-k}+\omep{p-k}} 
\Bigr\}
\Bigl\{
\frac{1}{2\omep{q-k}+\omep{p-k} + \omep{p}}  
\Bigr\}
\nonumber\\
& &\cdot \Bigl\{
\frac{1}{2\omep{q-k}+2\omep{p-k} + \omep{p}+ \omep{q}} 
\Bigr\}
\sum_{\nu\tau}V_{1\nu}(2p-k)T_+(p-k)V_{1\rho}(p+q-2k)
\nonumber \\
& &\cdot T_+(q-k)
V_{1\tau}(2q-k)D_{\nu\tau}(k) + (5\,\,\,terms).
\label{eqn:gamma}
\ee 
As is seen, these terms are described by the Feynman diagram Fig. 2. (c).  
In the limit $a \rightarrow 0$, $\Gamma_\rho(p,q) $ is evaluated in the same 
way to $\Sigma(p)$. First we perform the analytic continuation 
(\ref{eqn:dim}) and the rescaling of the loop momentum $k \rightarrow 
{\bar k}=ka$. Then we expand $\Gamma_\rho(p,q)$ with respect to 
the external momentums $p$ and $q$ to extract the divergence in the limit 
$a \rightarrow 0$. In the present 
case, only the zero term of the Taylor expansion leads to the logarithmic 
divergence, which appears as the infrared divergence in the 
${\bar k}$ integration. The infrared divergence takes place only 
in the vicinity ${\bar k} \simeq 0$, because of the pole structure 
of the propagators, and is evaluated by expanding the integrand around 
${\bar k} = 0$. Therefore, all the $\omega_+$ factors are 
(after the rescaling) reduced to the constant $\lambda$ in this calculation. 
Summing up all the six terms, the result of the calculation is, 
\be 
\Gamma_\rho(p=0,q=0) &=& \frac{1}{8} (ig)^3 \frac{1}{2}(1-\gfive T_c) 
\frac{1}{16\pi^2} \gamma_\rho \{1-(1-\xi)\} 
\{ \frac{1}{\varepsilon} +\log (\kappa a)^2 \} 
\nonumber \\ 
+ finite \,\,\,terms.
\ee
The finite part is also evaluated in the same way as 
the self-energy case, and is given by, 
\be
\Gamma_\rho(p,q) -\Gamma_\rho(0,0)&=& 
-\frac{1}{8} (ig)^3 \frac{1}{2}(1-\gfive T_c) 
\frac{1}{16\pi^2} \gamma_\rho \{1-(1-\xi)\} 
\{ \frac{1}{\varepsilon} +\log (\kappa / \mu)^2
\nonumber \\
& & + p,q,\mu \,\,\,dependent \,\,\,terms  \},
\ee 
where we introduced the renormalization point $\mu$. 
$\Gamma_\rho(p,q) $ is the sum of the above two terms and is independent of 
$\kappa$. Similarly, 
\be
& &\svpd  G_+ {\cal V}  G_+ {\cal V} \no G_- {\cal V} \svm \rightarrow 
(2\pi)^4 \delta_P (p+t-q) \sum_{\rho}D_{\mu\rho}(t) 
e^{ -\frac{i}{2}(p+t-q)_\rho a } 
\nonumber \\
& &
\cdot \Bigl\{
S_-(p) \frac{a}{\lambda} \Gamma^1_\rho(p,q) S_+(q) 
-
S_+(p) \frac{a}{\lambda} \Gamma^2_\rho(p,q) S_+(q) 
-
S_-(p) \frac{a}{\lambda} \Gamma^3_\rho(p,q) S_-(q) 
\Bigr\}, 
\ee 
where $\Gamma^{1,2,3}_\rho(p,q) $ are evaluated in the same way 
and they are given by (in the limit $a \rightarrow 0$), 
\be
\Gamma^{1,2,3}_\rho(p,q) \rightarrow 
\frac{1}{8} (ig)^3 \frac{1}{2}(1-\gfive T_c) 
\frac{1}{16\pi^2} \gamma_\rho \{1-(1-\xi)\} \log (a \mu)^2
 + finite \,\,\,terms. 
\ee
The other two terms, $\svpd {\cal V} G_+ \no G_- {\cal V} G_- {\cal V} \svm$ 
and $\svpd \no G_- {\cal V} G_- {\cal V} G_- {\cal V} \svm$, 
in eq. (\ref{eqn:totv}) 
are also described by the Feynman diagrams Fig. 2. (c) 
and yield similar results. 
Thus, all the divergent terms respect the correct chiral property of the 
vertex. Summing up all the contributions, 
the renormalization factor of the vertex is, 
\be
Z_1 &=& 1 + \{\frac{1}{8} + \frac{3}{8}+ \frac{3}{8} + \frac{1}{8} \}  
\frac{g^2}{16\pi^2} (\log (a \mu)^2 + const )
\nonumber \\
&=& Z_2,
\ee 
where $1/8$, $3/8$, $3/8$ 
and $1/8$ indicate the contributions of the first, second, third and fourth 
terms in eq. (\ref{eqn:logv}), respectively. 
This proves that the vertex is correctly renormalized.

\section{Discussion}

\indent

We have confirmed that, taking into account the dynamical nature of the gauge 
fields, the fermion propagator and fermion-fermion-gauge boson vertex are 
properly renormalized preserving the correct chiral properties. 
Our analysis, together with the studies of the 
vacuum polarization \cite{vac} and the gauge boson n-point functions 
\cite{n}, proves the renormalizability of a chiral gauge theory on the 
lattice in the overlap formulation at one loop level. 

Within the one loop level, phase conventions of the states 
$\svapm$, namely the imaginary parts of $\alpha_\pm(A)$ do not affect 
our calculations. For higher orders, however, the phase conventions of 
$\alpha_\pm(A)$, which, in some case, might not be consistent with the gauge 
invariance, will enter into the analysis. 
Better understanding of this point, even though within the 
perturbation theory, will provide qualitatively new informations on 
this formulation.


\section*{Acknowledgment}

\indent

We would like to thank Y.~Kikukawa, R.~Narayanan, 
S.~Randjbar-Daemi and G.~Thompson for useful discussions, and 
M.~O'Loughlin for useful comments on the manuscript.


\section*{Figure Caption}
\renewcommand{\labelenumi}{Fig.~\arabic{enumi}}
\begin{enumerate}
\item 
The Feynman diagrams describing the self-energy corrections. 
The diagram (a) yields the linearly and logarithmically divergent 
terms, while the diagram (b) yields only the linearly divergent 
terms. The linearly divergent terms in these diagrams are reduced to 
finite wave function renormalization factors at the level of the 
propagator. 
\item 
The Feynman diagrams describing the vertex corrections. 
The diagrams (a) and (b) lead only to the finite 
corrections. The diagram (c), containing only the interaction ${\cal V}_1$, 
leads to the logarithmic divergences. 
\end{enumerate}

\section*{Figures}
\input FEYNMAN

\begin{picture}(8000,8000)
\drawline\fermion[\E\REG](0,0)[2000]
\drawloop\gluon[\NE 3](\pbackx,\pbacky)
\drawline\fermion[\E\REG](\pbackx,\pbacky)[2000]
\drawline\fermion[\W\REG](\pbackx,\pbacky)[7000]
\end{picture}
\hspace{3cm}
\begin{picture}(8000,8000)
\drawline\fermion[\E\REG](0,0)[5000]
\drawloop\gluon[\W 8](\pbackx,\pbacky)
\drawline\fermion[\E\REG](\pbackx,\pbacky)[4000]
\drawline\fermion[\W\REG](\pbackx,\pbacky)[7000]
\end{picture}

\vspace{.5cm}

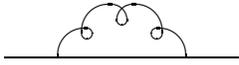
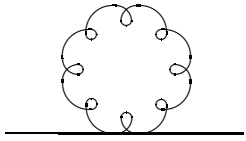
Fig. 1. (a). \hspace{4cm} Fig. 1. (b).

\vspace{.5cm}

\begin{picture}(8000,8000)
\drawline\fermion[\E\REG](0,0)[2000]
\drawloop\gluon[\NE 3](\pbackx,\pbacky)
\drawline\fermion[\E\REG](\pbackx,\pbacky)[2000]
\drawline\fermion[\W\REG](\pbackx,\pbacky)[7000]
\drawline\gluon[\S\CENTRAL](\pbackx,\pbacky)[4]
\end{picture}
\hspace{3cm}
\begin{picture}(8000,8000)
\drawline\fermion[\E\REG](0,0)[2000]
\drawloop\gluon[\NE 3](\pbackx,\pbacky)
\drawline\fermion[\E\REG](\pbackx,\pbacky)[2000]
\drawline\fermion[\W\REG](\pbackx,\pbacky)[7000]
\drawline\gluon[\S\REG](\gluonbackx,\pfronty)[4]
\end{picture}

\vspace{1cm}

\hspace{4cm} Figs. 2. (a). 

\vspace{.5cm}

\begin{picture}(8000,8000)
\drawline\fermion[\E\REG](0,0)[5000]
\drawloop\gluon[\W 8](\pbackx,\pbacky)
\drawline\fermion[\E\REG](\pbackx,\pbacky)[4000]
\drawline\fermion[\W\REG](\pbackx,\pbacky)[7000]
\drawline\gluon[\S \REG](\loopmidx,\pbacky)[4]
\end{picture}
\hspace{3cm}
\begin{picture}(8000,8000)
\drawline\fermion[\E\REG](0,0)[2000]
\drawloop\gluon[\NE 3](\pbackx,\pbacky)
\drawline\fermion[\E\REG](\pbackx,\pbacky)[2000]
\drawline\fermion[\W\REG](\pbackx,\pbacky)[7000]
\drawline\gluon[\S\CENTRAL](\loopbackx,\pfronty)[4]
\end{picture}

\vspace{2cm}

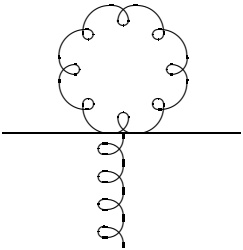
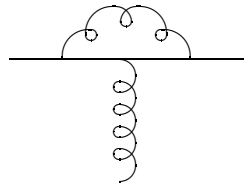
Fig. 2. (b). \hspace{4cm} Fig. 2. (c).


\begin{thebibliography}{99}
\bibitem{gam}
The Euclidean gamma matrices $ \gamma^{E}_{\mu}$ are related to the 
Minkowskian gamma matrices $ \gamma^{\mu}_{M}$ as 
$\gamma^{E}_{i}= -i \gamma^{i}_{M}$ (for $i=1,2,3$) and 
$\gamma^{E}_{4}= \gamma^{0}_{M}$.
\bibitem{phase}
Nontrivial phases of 
$\alpha_{\pm}(A) $ appear at the order ${\cal O}(g^3)$ \cite{n,kikukawa} 
and do not affect the calculations of the propagator and vertex 
at one loop level.
\bibitem{dimensional}
To be specific, we understand that the regularization (\ref{eqn:dim}) 
is performed after the manipulation of the gamma matrices 
in $\Sigma(p) $ even though we introduce from the beginning 
before the calculation of the gamma matrices. In this sense, 
the regularization (\ref{eqn:dim}) is similar to the $\bar{DR}$ 
regularization of the ultraviolet divergence.
\end{thebibliography}
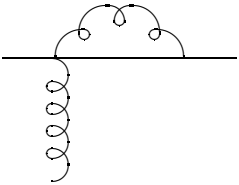
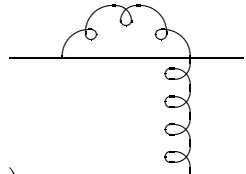
\end{document}